\documentclass[aps,preprint]{revtex4}
\usepackage{pifont}
\usepackage{mathrsfs}
\usepackage{amsfonts}
\usepackage{amsmath}
\usepackage{amssymb}
\usepackage{graphicx}%
\begin{document}
\title{Bohmian mechanics to high-order harmonic generation}
\author{X. Y. Lai$^{1,2}$}
\author{Qingyu Cai$^{1}$}
\email{qycai@wipm.ac.cn}
\author{M. S. Zhan$^{1,3}$}
\affiliation{$^{1}$State Key Laboratory of Magnetic Resonances and Atomic and Molecular
Physics, Wuhan Institute of Physics and Mathematics, The Chinese Academy of
Sciences, Wuhan 430071, China}
\affiliation{$^{2}$Graduation University of Chinese Academy of Sciences, Beijing 100081, China}
\affiliation{$^{3}$Center for Cold Atom Physics, The Chinese Academy of Sciences, Wuhan
430071, China}

\begin{abstract}
This paper introduces Bohmian mechanics (BM) into the intense
laser-atom physics to study high-order harmonic generation. In BM,
the trajectories of atomic electrons in intense laser field can be
obtained with the Bohm-Newton equation. The power spectrum with the
trajectory of an atomic electron is calculated, which is found to be
irregular. Next, the power spectrum associated with an atom ensemble
from BM is considered, where the power spectrum becomes regular and
consistent with that from quantum mechanics. Finally, the reason of
the generation of the irregular spectrum is discussed.
\end{abstract}

\pacs{PACC numbers: 0365, 3280K}

\maketitle

\section{Introduction}
Atom interacting with the intense laser field (ILF) can absorb more
than one photon and then emits photons at harmonics of the laser
frequency, which is well known as high-order harmonic generation
(HHG) \cite{Burnett1,ma,Ferray}. The harmonic spectrum has the
following characteristics as a function of increasing frequency: a
rapid decrease in harmonic intensity at first, followed by a plateau
region with the harmonics having similar intensities; and then a
rapid drop in harmonic strength beyond the plateau region. Recently,
lots of theoretical works have studied this multiphoton phenomenon,
including solving the time-dependent Schr\"{o}dinger equation
\cite{Burnett1,Zhou,Qiao,bian,zzy,zzy2}, the semiclassical
trajectory methods, such as the three-step model \cite{Corkum} and
the Feynman's path-integral approach in the strong field
approximation \cite{Salieres,Salieres2,eden}, and even the classical
trajectory method \cite{gb}.

Bohmian mechanics (BM) \cite{Holland,bohm,Nikolic}, or called
quantum trajectory method \cite{rew}, is another alternative
formalism of quantum mechanics. In the early years, BM has been used
to study some fundamental quantum phenomena such as the scattering
by a square potential barrier \cite{joh,cd} and the diffraction
through two slits \cite{cp}. Recently, it has been applied to study
and analyze many different processes and phenomena in physics and
chemistry, such as atom-surface scattering \cite{ass3,ass}, electron
transport in mesoscopic systems \cite{xo}, photodissociation of NOCl
and NO$_{2}$ \cite{bkd}, and the chemical reactions \cite{rew2}.
Also, BM in terms of quantum trajectory has been considered to study
chaos \cite{dd,us,mhp,ce}. Recently, it has been successfully
applied to the intense laser-atom physics to study the dynamics of
the above-threshold ionization (ATI) photoelectron by authors
\cite{lcz}.

In this paper, BM is introduced to the intense laser-atom physics to
study HHG, which allows us to follow the time evolution of
individual electron trajectory in atomic system first. We then
calculate the power spectrum with the electron trajectory and find
the corresponding spectrum is irregular. Next, we consider the power
spectrum associated with an atom ensemble. In the atom ensemble,
harmonic generation is determined by the Maxwell equation, where the
electronic polarization of the atom ensemble plays a key role for
the harmonic generation. We will show the electronic polarization of
the atom ensemble gained from BM is equivalent to that obtained from
quantum mechanics. On this condition, the power spectrum from BM is
regular and consistent with that obtained from quantum mechanics.
Finally, we will briefly discuss the reason why an individual
electron trajectory generates the irregular spectrum.

This paper is organized as follow: We will briefly introduce quantum
trajectory method first. Then we show the Hamiltonian for the
hydrogen atom in ILF. In Sec. IV, we calculate the harmonic spectrum
from a single electron trajectory, following by harmonic generation
associated with an ensemble of atoms interacting with the ILF in
Sec. V. Finally, we briefly discuss the reason of the generation of
the irregular spectrum with an individual electron trajectory and
then conclude.

\section{Quantum trajectory formalism of Bohmian mechanics}

Quantum trajectory method comes from the following transformation of
the time-dependent Schr\"{o}dinger equation \cite{Holland,bohm}.
Firstly a wave function can be written in the polar form
$\psi(\mathbf{r},t)=R(\mathbf{r},t)\exp (iS(\mathbf{r},t)/\hbar)$.
Secondly $\psi(\mathbf{r},t)$ is substituted into the
time-dependent Schr\"{o}dinger equation. Then the real part of the resulting equation is%
\begin{equation}
\frac{\partial S}{\partial t}+\frac{\left(  \nabla S\right)
^{2}}{2m}+V+Q=0
\label{H1}%
\end{equation}
and the imaginary part has the form
\begin{equation}
\frac{\partial\rho}{\partial t}+\nabla(\rho\mathbf{v})=0, \label{H2}%
\end{equation}
where $\rho(\mathbf{r},t)=R^{2}(\mathbf{r},t)$, $\mathbf{v}=\nabla
S(\mathbf{r},t)/m$, and
$Q(\mathbf{r},t)=-\frac{\hbar^{2}}{2m}\frac{\nabla ^{2}R}{R}$. These
two equations look like the classical Hamilton-Jacobi equation and
the equation of continuity, respectively, except Eq. (\ref{H1}) has
an extra term $Q$ which is usually called quantum potential in BM.
Thus Eq. (\ref{H1}) is called the quantum Hamilton-Jacobi equation
governed by an external potential $V$ and the quantum potential $Q$.
The motion of
particle is guided by the Bohm-Newton equation of motion, $md^{2}\mathbf{r}%
/dt^{2}=-\nabla(V+Q)$, or, equivalently, by
\begin{equation}
d\mathbf{r}/dt=\nabla S(\mathbf{r},t)/m. \label{eq}%
\end{equation}
In practice, we solve the time-dependent Schr\"{o}dinger equation to
obtain $\psi(\mathbf{r},t)$, and then $S(\mathbf{r},t)$. In this
way, the quantum trajectory of the particle can be gained by
integrating Eq. (\ref{eq}).

\section{Numerical solution of the time-dependent Schr\"{o}dinger equation}

The Schr\"{o}dinger equation for Hydrogen atom in ILF can be written
as (atomic units are used throughout)
\[
i\frac{\partial\psi(\mathbf{r},t)}{\partial t}=[\widehat{H}_{0}(\mathbf{r}%
)+\widehat{V}(\mathbf{r},t)]\psi(\mathbf{r},t).
\]
Here $\widehat{H}_{0}(\mathbf{r})$ is the field-free Hydrogen atom
Hamiltonian and $\widehat{V}(\mathbf{r},t)$ is the intense
laser-atom interaction:
$\widehat{H}_{0}(\mathbf{r})=-\frac{1}{2}\frac{d^{2}}{dr^{2}}+\frac
{\widehat{L}^{2}}{2r^{2}}-\frac{1}{r},$ $\widehat{V}(\mathbf{r}%
,t)=-\mathbf{r\cdot E}(t)=-zE(t),$ where the laser field is the
linearly polarized field ($\mathbf{E}||\mathbf{z}$) and $E(t)$ is
the laser field profile. Due to the linearly polarized laser field,
magnetic quantum number $m$ of Hydrogen atom is a good quantum
number, so that the problem of solving the time-dependent
Schr\"{o}dinger here can be simplified into a two-dimensional
problem (see the inset in Fig. 1).
In this work the solution of the time-dependent Schr\"{o}dinger
equation $\psi(\mathbf{r},t)$ is obtained by the grid method and the
second-order split-operator method, which has been detailedly
introduced by Tong \textit{et al} \cite{xmt}. The laser field
profile is $E(t)=\left\{
  \begin{array}{ll}
    E_{0}\sin^{2}(\frac{\pi t}{20T})\sin(\omega_{0}t), & {0\leq t\leq10T;} \\
    E_{0}\sin(\omega_{0}t), & {t>10T,}
  \end{array}
\right.$ where $T=2\pi/\omega_{0}$, and $E_{0}$ and $\omega_{0}$ are
the electric field amplitude and angular frequency, respectively. We
choose $E_{0}=0.0292$ a.u. (i.e., $I=3.0\times10^{13}$W/cm$^{2}$),
and $\omega_{0}=0.0587$ a.u. (i.e., $ \lambda=775$nm). The initial
wave function $\psi(\mathbf{r},0)$ of the system is in the ground
state of the field-free Hydrogen atom.

\begin{figure*}[ptbh]
\centering \label{1}
\includegraphics[width=3.3in]{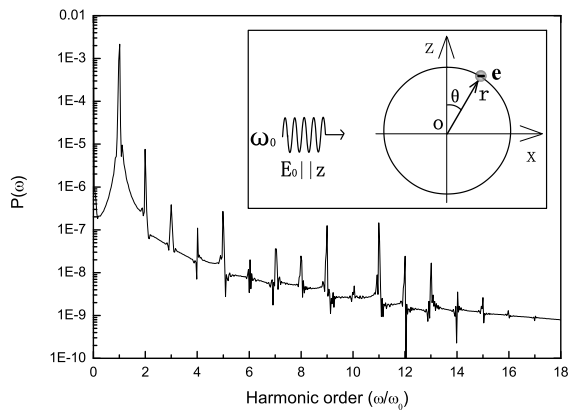}\caption{Power spectrum of the electron with the
initial position $\mathbf{r}_{0}=(1.0$ a.u., $\pi/4)$. The inset
shows the direction of the laser beam and the electron coordinate of
atomic system.}%
\label{fig2}%
\end{figure*}

\section{Harmonic spectrum associated with a single electron trajectory}

After obtaining the wavefunction $\psi(\mathbf{r},t)$, we can follow
the time evolution of electron trajectory $\mathbf{r}(t)$
 by integrating Eq. (\ref{eq}) with the electron
initial position $\mathbf{r}_{0}$. According to BM
\cite{Holland,bohm,Nikolic}, the initial electron distribution is
$|\psi(\mathbf{r},0)|^{2}$. In this work, $\psi(\mathbf{r},0)$ is
the ground state of the field-free Hydrogen atom. Thus we can obtain
an ensemble of electron trajectories with the corresponding electron
initial positions. For an individual electron trajectory, the atomic
dipole moment $\mu_{z}(t)$ along the laser polarization direction is
$\mathbf{\hat{z}}\cdot\mathbf{r}(t)$. The corresponding power
spectrum is gained by the Fourier transformation of the dipole
moment \cite{gb,kck}:
\begin{equation}
P(\omega)=\left|\frac{1}{t_{f}-t_{i}}\int_{t_{i}}^{t_{f}}\mu_{z}(t)e^{-i\omega
t}\right|^{2},
 \label{fd}
\end{equation}
where $t_{i}=0$ and the propagation time $t_{f}$ is $35T$ in this
work.

Explicitly, we take one atomic electron as an example with the
initial position $\mathbf{r}_{0}=(1.0$ a.u., $\pi/4)$ in polar
coordinates. The electron trajectory $\mathbf{r}(t)$ is calculated
 by integrating Eq. (\ref{eq}). The corresponding power spectrum is shown in Fig. 1 from Eq.
(\ref{fd}). The spectrum has even harmonics, which is dominated by a
Rayleigh scattering component at the laser frequency
$\omega=\omega_{0}$. We have obtained lots of electron trajectories
with different initial position $\mathbf{r}_{0}$ according to the
initial electron distribution $|\psi (\mathbf{r},0)|^{2}$. The
corresponding power spectra have the similar character described
above. In the following, however, we will show the even harmonics of
the power spectrum are coherently removed in an ensemble of atoms
and the corresponding power spectrum is consistent with that
obtained from quantum mechanics.
\section{Harmonic generation from an ensemble of atoms interacting with the
intense laser field}

The harmonic generation of an atom ensemble can be gained with the
Maxwell equation  \cite{al1,al2,al3,ya}:%
\[
\nabla^{2}\mathscr{E}-\frac{1}{c^{2}}\frac{\partial^{2}\mathscr{E}}{\partial
t^{2}}=\frac{4\pi}{c^{2}}\frac{\partial^{2}\mathscr{P}}{\partial
t^{2}},
\]
where $\mathscr{P}(\mathbf{r}^{\prime},t)$ is the electronic
polarization of the atom  ensemble at the point
$\mathbf{r}^{\prime}$ and $\mathscr{E}(\mathbf{r}^{\prime},t)$
describes the electromagnetic field (involving the fundamental and
the harmonic fields). Obviously, the electronic polarization
$\mathscr{P}(\mathbf{r}^{\prime},t)$ plays a key role for the
harmonic generation from the ensemble of atoms. We will show that
the electronic polarization $\mathscr{P}(\mathbf{r}^{\prime},t)$
gained from BM is, approximately, equivalent to that obtained from
quantum mechanics, i.e., both of their harmonic spectra from the
atom ensemble are consistent.

We first assume the number of atoms is $N$ in a small volume $\Delta
V$ near the point $\mathbf{r}^{\prime}$ in the atom ensemble (
$\Delta V \ll d^{3}$, where $d$ is the focal spot size of laser).
The electronic polarization $\mathscr{P}(\mathbf{r}^{\prime})$ is
defined as the atomic dipole moment per unit volume:
$\mathscr{P}(\mathbf{r}^{\prime})=
\sum\limits_{i}^{N}\mathbf{p}_{i}\mathbf{/}\Delta V$, where
$\mathbf{p}_{i}$ is the atomic dipole moment of each atom in the
small  volume $\Delta V$. From the view point of quantum mechanics,
the
dipole moment of each atom is $p(t)=\left\langle \psi(\mathbf{r}%
,t)|z|\psi(\mathbf{r},t)\right\rangle $ in the $\mathbf{\hat{z}}$
direction, where $\mathbf{r}$ is the electron coordinate of atomic
system and $\psi(\mathbf{r},t)$ is the electron wavefunction. Thus
the electronic polarization of the atom ensemble obtained from
quantum mechanics is
\begin{equation}
\mathscr{P}(\mathbf{r}^{\prime},t)=N p(t)/\Delta V\label{s}%
\end{equation}
in the $\mathbf{\hat{z}}$ direction.

\begin{figure*}[ptbh]
\centering \label{2}
\includegraphics[width=3.3in]{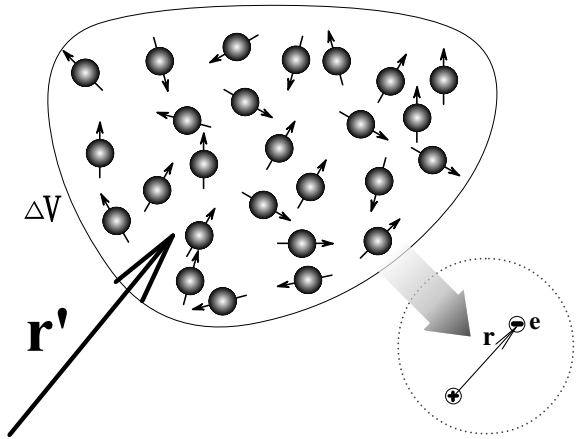}\caption{A small volume $\Delta V$ near the point $\mathbf{r}^{\prime}$ in the atom
ensemble, which includes lots of atoms with the different dipole
moments from the view of BM. The inset shows the electron coordinate
of atomic system.}%
\label{fig2}%
\end{figure*}

On the other hand, we calculate the electronic polarization of the
 atom ensemble from  BM. According to BM
\cite{Holland,bohm,Nikolic}, the probability that an electron lies
between $\mathbf{r}$ and $\mathbf{r}+\mathbf{dr}$ in the atomic
system
at the time $t$ is given by $|\psi(\mathbf{r},t)|^{2}\mathbf{dr}%
$, where the corresponding atomic dipole moment is $\mathbf{r}$ (in
atomic units). Then in the small volume $\Delta V$ of the atom
ensemble, the number of atoms with atomic dipole moment $\mathbf{r}$
is $(N|\psi(\mathbf{r},t)|^{2} \mathbf{dr})$. The sum of all kinds
of atomic dipole moments in the volume $\Delta V$ is
$\sum\limits_{j}N|\psi(\mathbf{r}_{j},t)|^{2}\mathbf{dr}_{j}\mathbf{r}%
_{j}$. Thus the electronic polarization
$\mathscr{P}(\mathbf{r}^{\prime},t)$ at the point
$\mathbf{r}^{\prime}$ in the atom ensemble is
$\sum\limits_{j}N|\psi(\mathbf{r}_{j},t)|^{2}\mathbf{dr}_{j}\mathbf{r}%
_{j}/\Delta V$ (see Fig. 2). For an atom ensemble, we can replace
the sum by an integral,
\begin{equation}
\sum\limits_{j}\frac{N|\psi(\mathbf{r}_{j},t)|^{2}\mathbf{dr}_{j}\mathbf{r}%
_{j}}{\Delta V} \rightarrow \frac{N}{\Delta V}\int|\psi(\mathbf{r},t)|^{2}\mathbf{r}d\mathbf{r}%
.\label{eq6}
\end{equation}
Further, because the laser field is the linearly polarized field
($\mathbf{E}||\mathbf{z}$), the time-dependent wavefunction
$\psi(\mathbf{r},t)$ is symmetrical along $\mathbf{\hat{z}}$ axis.
Thus in the $\mathbf{\hat{z}}$ direction,
\begin{align}
\frac{N}{\Delta V}\int|\psi(\mathbf{r},t)|^{2}
\mathbf{r}d\mathbf{r} %
  &= \frac{N}{\Delta V} \int|\psi(\mathbf{r},t)|^{2}r
\cos\theta d\mathbf{r} \nonumber %
\\ &= \frac{N}{\Delta V}\int|\psi(\mathbf{r},t)|^{2}zd\mathbf{r},%
  \label{eq7}
\end{align}
 where the last term equals Eq.
(\ref{s}). In this way, the electronic polarization
$\mathscr{P}(\mathbf{r}^{\prime},t)$ gained from BM is,
approximately, equivalent to that obtained from quantum mechanics,
i.e., both of their harmonic spectra from the atom ensemble are the
same. Figure 3 shows the harmonic spectrum from an ensemble of
Hydrogen atoms interacting with ILF (laser confocal parameter
$b=5.0$ mm, gas density $10^{17}$ atoms/cm$^{3}$ described by a
truncated Lorentzian in the $\mathbf{\hat{x}}$ direction with
$L=0.5$ mm \cite{al1}), which has a clear plateau region and cutoff
at the $11$th order.

\begin{figure*}[ptbh]
\centering \label{3}
\includegraphics[width=3.3in]{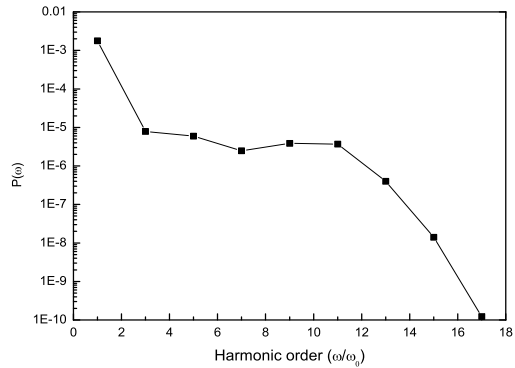}\caption{Harmonic spectrum associated with an ensemble of Hydrogen atoms at
the laser field $I=3.0\times10^{13}$W/cm$^{2}$ and $ \lambda=775
$nm.}%
\label{fig2}%
\end{figure*}

\section{Discussion}
In BM, harmonic spectrum from a single electron trajectory has even
harmonics (see Fig. 1), but they are removed in the spectrum
associated with an ensemble of atoms (see Fig. 3). The reason is
that an individual electron trajectory does not possess an inversion
center \cite{gb}, but the electron trajectories of the atom ensemble
have such centre. Here we will numerically show that two electron
trajectories are enough to get rid of the unphysical even harmonics
if the initial positions $\mathbf{r}_{0}$ of the two electrons are
centrally symmetric. Let's take two electrons as an example with the
initial positions $\mathbf{r}_{0}^{1}=(1.0$ a.u., $\pi/4)$ and
$\mathbf{r}_{0}^{2}=(1.0$ a.u., $5\pi/4)$, respectively. The
corresponding atomic dipole moment $\mu_{z}(t)$ along the laser
polarization direction is $\mathbf{\hat{z}}\cdot
[\mathbf{r}^{1}(t)+\mathbf{r}^{2}(t)]$ and the power spectrum gained
from Eq. (4) is shown in Fig. 4 (solid curve), which has only
odd-order harmonics. In addition, we calculate the HHG power
spectrum in the length form from the time-dependent Schr\"{o}dinger
equation (dotted curve in Fig. 4) \cite{xmt}. Note that these two
curves can  basically overlap  if the value of the power spectrum
from BM is multiplied by a factor of 0.4, i.e., both of them have
the same relative intensity. This is an interesting result, but the
reason why HHG power spectra from BM and the time-dependent
Schr\"{o}dinger equation have the same relative intensity should be
studied in the future.

\begin{figure*}[ptbh]
\centering \label{4}
\includegraphics[width=3.3in]{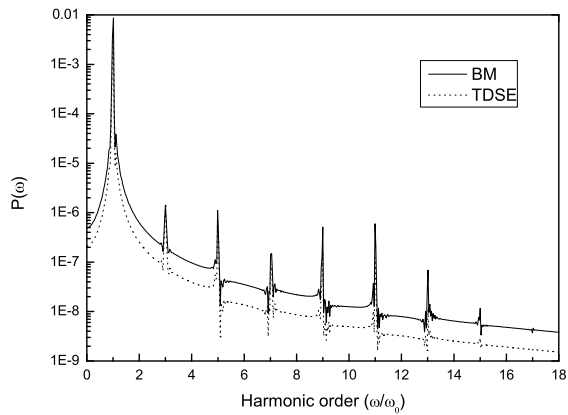}\caption{The solid curve corresponds to power spectrum of two Bohmian electrons with the centrally symmetric
initial positions: $\mathbf{r}_{0}^{1}=(1.0$ a.u., $\pi/4)$ and
$\mathbf{r}_{0}^{2}=(1.0$ a.u., $5\pi/4)$. The dot is the HHG power
spectrum in the length form from the time-dependent Schr\"{o}dinger
equation (TDSE).}%
\label{fig2}%
\end{figure*}

\section{Summary}

In summary, we introduce BM into the intense laser-atom physics to
discuss HHG. It allows us to follow the time evolution of each
electron trajectory in an atomic system. We find that the power
spectrum from an individual electron trajectory has the even
unphysical harmonics. But the even harmonics are coherently removed
in the ensemble of atoms and the power spectrum is consistent with
that obtained from quantum mechanics. The reason of the appearance
of the even harmonics is an individual electron trajectory does not
possess an inversion center, but the electron trajectories of the
atom ensemble have such centre.

\end{document}